\documentclass[useAMS]{mn2e}
\usepackage{graphicx}

\usepackage{times}

\title[First $\gamma$-ray outburst of the NLS1 PMN~J0948$+$0022]{The first gamma-ray outburst of a Narrow-Line Seyfert 1 Galaxy: the case of PMN~J0948$+$0022 in July~2010}
\author[L. Foschini et al.]{L. Foschini$^{1}$\thanks{E-mail: \texttt{luigi.foschini@brera.inaf.it}.}, 
G. Ghisellini$^{1}$, 
Y.~Y. Kovalev$^{2,3}$, 
M.~L. Lister$^{4}$, 
F. D'Ammando$^{5}$\thanks{\emph{Fermi} LAT Member/Affiliated}, \newauthor
D.~J. Thompson$^{6}$\footnotemark[2], 
A. Tramacere$^{7}$\footnotemark[2], 
E. Angelakis$^{3}$,
D. Donato$^{8,9}$\footnotemark[2], 
A. Falcone$^{10}$, \newauthor 
L. Fuhrmann$^{3}$\footnotemark[2], 
M. Hauser$^{11}$, 
Yu.~A. Kovalev$^{2}$, 
K. Mannheim$^{12}$, 
L. Maraschi$^{1}$, \newauthor 
W. Max-Moerbeck$^{13}$\footnotemark[2], 
I. Nestoras$^{3}$\footnotemark[2], 
V. Pavlidou$^{13}$\footnotemark[2],
T.~J. Pearson$^{13}$\footnotemark[2],\newauthor
A.~B. Pushkarev$^{14,15,3}$,
A.~C.~S. Readhead$^{13}$\footnotemark[2], 
J.~L. Richards$^{13}$\footnotemark[2],
M.~A. Stevenson$^{13}$\footnotemark[2],\newauthor
G. Tagliaferri$^{1}$, 
O. Tibolla$^{12}$\footnotemark[2], 
F. Tavecchio$^{1}$, 
S. Wagner$^{11}$\\
$^{1}$ INAF - Osservatorio Astronomico di Brera, via E. Bianchi 46, 23807, Merate (LC), Italy\\
$^{2}$ Astro Space Center of the Lebedev Physical Institute, Profsoyuznaya 84/32, 117997 Moscow, Russia\\
$^{3}$ Max-Planck-Institut f\"ur Radioastronomie, Auf dem H\"ugel 69, D-53121 Bonn, Germany\\
$^{4}$ Department of Physics, Purdue University, 525 Northwestern Avenue, West Lafayette, IN 47907, USA\\
$^{5}$ INAF -- IASF-Palermo, Via Ugo La Malfa 153, 90146 Palermo, Italy\\
$^{6}$ NASA Goddard Space Flight Center, Greenbelt, MD 20771, USA\\
$^{7}$ ISDC Data Centre for Astrophysics, Chemin d'Ecogia 16, CH-1290, Versoix, Switzerland\\
$^{8}$ Center for Research and Exploration in Space Science and Technology and NASA Goddard Space Flight Center, Greenbelt, MD 20771, USA.\\
$^{9}$ Department of Physics and Department of Astronomy, University of Maryland, College Park, MD 20742, USA.\\
$^{10}$ Department of Astronomy \& Astrophysics, Pennsylvania State University, University Park, PA 16802, USA\\
$^{11}$ Landessternwarte, Universit\"at Heidelberg, K\"onigstuhl, D 69117 Heidelberg, Germany\\
$^{12}$ University of W\"urzburg, 97074, W\"urzburg, Germany\\
$^{13}$ Cahill Center for Astronomy and Astrophysics, California Institute of Technology, Pasadena, CA 91125, USA\\
$^{14}$ Crimean Astrophysical Observatory, 98409 Nauchny, Crimea, Ukraine\\
$^{15}$ Pulkovo Observatory, 196140 St. Petersburg, Russia
}

\begin{document}

\date{Accepted 2010 December 20.  Received 2010 December 3; in original form 2010 October 21}

\pagerange{\pageref{firstpage}--\pageref{lastpage}} \pubyear{2010}

\maketitle

\label{firstpage}

\begin{abstract}
We report on a multiwavelength campaign on the radio-loud Narrow-Line Seyfert 1 (NLS1) Galaxy PMN~J0948$+$0022 ($z = 0.5846$) performed in 2010~July-September and triggered by a high-energy $\gamma$-ray outburst observed by the Large Area Telescope (LAT) onboard the \emph{Fermi Gamma-ray Space Telescope}. The peak flux in the 0.1$-$100~GeV energy band exceeded, for the first time in this type of source, the value of $\sim 10^{-6}$~ph~cm$^{-2}$~s$^{-1}$, corresponding to an observed luminosity of $\sim 10^{48}$~erg~s$^{-1}$. Although the source was too close to the Sun position to organize a densely sampled follow-up, it was possible to gather some multiwavelength data that confirmed the state of high activity across the sampled electromagnetic spectrum. The comparison of the spectral energy distribution of the NLS1 PMN~J0948$+$0022 with that of a typical blazar -- like 3C~273 -- shows that the power emitted at $\gamma$ rays is extreme. 
\end{abstract}


\section{Introduction}
Relativistic jets are the most extreme expression of the power that can be generated by supermassive black holes at the centres of galaxies. Their bolometric luminosity can reach values up to $\sim 10^{49-50}$~erg~s$^{-1}$, peaking in the $\gamma$-ray (MeV-TeV) energy band. To date, two classes of active galactic nuclei (AGNs) are known to generate these structures: blazars and radio galaxies, both hosted in giant elliptical galaxies (Blandford \& Rees 1978). The first class is composed of AGNs with the jet viewed at small angles and, therefore, the effects of the special relativity play a dominant role in shaping the emission of the electromagnetic radiation. In the second class -- radio galaxies -- the jets are viewed at larger angles and therefore the Doppler boosting is less intense. 

\begin{figure*}
\begin{center}
 \includegraphics[angle=270,scale=0.5]{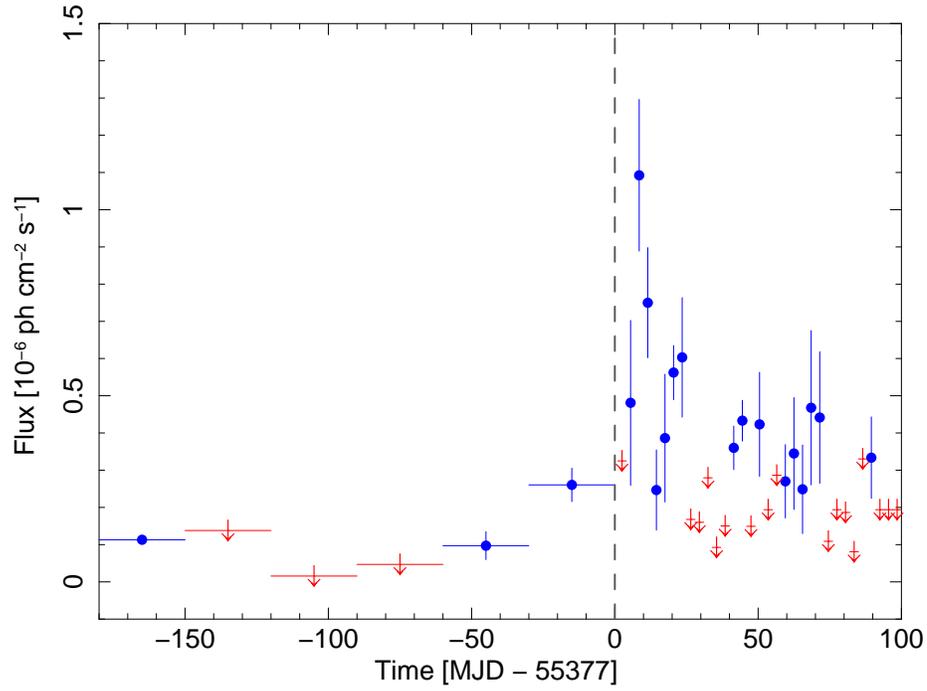} 
 \caption{\emph{Fermi}/LAT lightcurve (0.1$-$100~GeV) of PMN~J0948$+$0022 in 2010, binned over 1 month during the first six months of 2010 and with $3$~days time bin for the period 2010~July-September. The value 0 on the abscissa corresponds to 2010~June~30 (MJD 55377). Blue points are detection with Test Statistic $TS>10$ (equivalent to $\sim 3\sigma$, see Mattox et al. 1996 for the definition), while the others are $2\sigma$ upper limits. The abscissa has the same scale as Fig.~\ref{fig2}, to make comparison easier.}
   \label{fig1}
\end{center}
\end{figure*}

\begin{figure*}
\begin{center}
 \includegraphics[angle=270,scale=0.5]{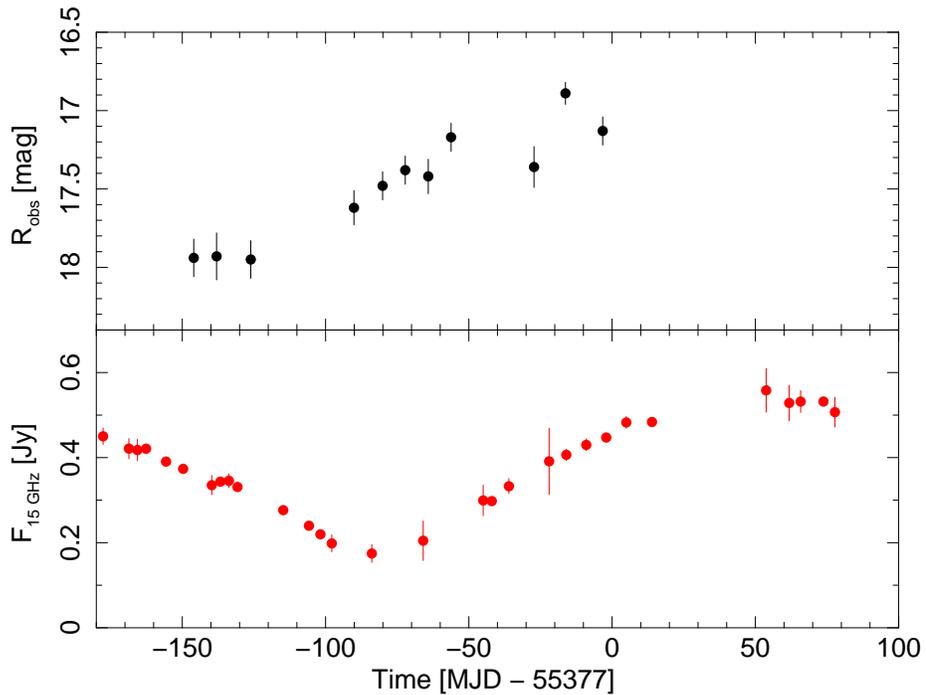} 
 \caption{Lightcurve of PMN~J0948$+$0022 at optical (ATOM, R filter, \emph{top panel}) and radio frequency (OVRO, 15~GHz, \emph{bottom panel}). The abscissa has the same scale as Fig.~\ref{fig1}, to make comparison easier. The value 0 corresponds to 2010~June~30 (MJD 55377).}
   \label{fig2}
\end{center}
\end{figure*}

The recent discovery of variable $\gamma$-ray emission from 4 NLS1s confirmed the presence of a third class of AGNs with relativistic jets (Abdo et al. 2009a,b,c, Foschini et al. 2010a), which was already suggested by the detection of variable radio emission with flat or inverted spectrum (e.g. Grupe et al. 2000, Komossa et al. 2006, Doi et al. 2006, Yuan et al. 2008). This finding poses intriguing questions about jet systems and how these structures are generated. One of these sources, PMN~J0948$+$0022 ($z=0.5846$) is classified as a typical NLS1, i.e. FWHM(H$\beta$)$ < 2000$~km~s$^{-1}$, [OIII]/H$\beta< 3$, with the presence of the FeII bump (see Pogge 2000 for a review on the characteristics of NLS1s). However, it displays also strong, compact and variable radio emission, with inverted spectrum, suggesting the presence of a relativistic jet (Zhou et al. 2003). The confirmation came with the detection of high-energy variable $\gamma$ rays by \emph{Fermi}/LAT (Abdo et al. 2009b, hereafter ``Discovery 2008''; Foschini et al. 2010a). A multiwavelength (MW) campaign performed in 2009~March-July displayed coordinated variability at all frequencies, thus confirming that the source detected by \emph{Fermi}/LAT is indeed the high-energy counterpart of PMN~J0948$+$0022 (Abdo et al. 2009c, hereafter ``MW Campaign 2009''). However, the variability observed in 2009 was much smaller in amplitude (more than a factor 2 at $\gamma$ rays) with respect to the 2010 campaign.

The main differences between NLS1s and the other AGNs with jets are the optical spectrum and the radio morphology, which is quite compact and without extended structures (Doi et al. 2006, Komossa et al. 2006, Abdo et al. 2009c, Foschini et al. 2010a, Gu \& Chen 2010). In addition, NLS1s are generally hosted in spiral galaxies (Deo et al. 2006, Zhou et al. 2006), which is at odds with the current observational paradigm linking very powerful jets with ellipticals (cf, e.g., Marscher 2009; see also B\"ottcher \& Dermer 2002). Obviously, to make a proper comparison we would need either a direct imaging of the host galaxies of the four $\gamma$-NLS1s or the morphologies of a similar sample in term of redshift and magnitude distribution. Our sources have redshifts between 0.061 and 0.585, while the work by Deo et al. (2006) is dedicated to a sample with $z<0.03$ and Zhou et al. (2006) studies the morphology of NLS1s with $z<0.1$, so there is only a partial overlap. At least one of the four NLS1s detected at $\gamma$ rays is definitely hosted by a spiral galaxy (1H~0323$+$342, see Zhou et al. 2007, Ant\'on et al. 2008), while for the others there are no high-resolution observations available. Therefore, some halo of doubt remains. 

Despite these uncertainties -- worth studying -- there are other novelties introduced by the $\gamma$-ray detection of NLS1s. The spectral energy distributions (SEDs) of the four $\gamma$-NLS1s point to AGNs with masses of the central black hole in the range of $10^{6-8} M_{\odot}$ and high accretion rates (up to 90\% of the Eddington value). These values are common for NLS1s (e.g. Boller et al. 1996, Grupe 2004, Collin \& Kawaguchi 2004), but not for blazars or radio galaxies. 

Another piece of the puzzle is given by the search for the parent population. Indeed, if $\gamma$-NLS1s seem to be similar to blazars, i.e. with the relativistic jet viewed at small angles, there should be a parent population with the jet viewed at large angles (as in the case of blazars vs radio galaxies). The first source of this type -- PKS~0558$-$504 ($z=0.137$) -- has been recently found by Gliozzi et al. (2010). Therefore, it seems that NLS1s could be a set of low mass systems ``parallel'' to blazars and radio galaxies. 

A key unknown in this research field was the power released by jets of $\gamma$-NLS1s. During the early observations and the 2009 MW campaign the maximum measured luminosity of PMN~J0948$+$0022, the most powerful of these NLS1s, was $\sim 10^{47}$~erg~s$^{-1}$ (0.1$-$100~GeV). On the other hand, it is known that blazars can reach greater luminosities (see, for example, Ghisellini et al. 2010), up to $\sim 10^{49}$~erg~s$^{-1}$ in the case of the brightest outbursts of 3C~454.3 (Bonnoli et al. 2011, Foschini et al. 2010b) or even in excess of $10^{50}$~erg~s$^{-1}$ in the recent episode of the gravitationally lensed quasar PKS~1830$-$21 (Ciprini et al. 2010). The question of the power could be studied in 2010~July, when PMN~J0948$+$0022 underwent a period of intense activity (Donato et al. 2010, Foschini 2010) with a peak flux of $\sim 10^{-6}$~ph~cm$^{-2}$~s$^{-1}$ (0.1$-$100~GeV), corresponding to an observed luminosity of $\sim 10^{48}$~erg~s$^{-1}$. Here we report about the study of this period. Some very preliminary results have been presented in Foschini et al. (2010c). 

In the following, we adopted a $\Lambda$CDM cosmology as calculated in Komatsu et al. (2009) based on \emph{WMAP} results: $h = 0.71$, $\Omega_m = 0.27$, $\Omega_\Lambda = 0.73$ and with the Hubble-Lema\^{i}tre constant $H_0=100h=71$~km~s$^{-1}$~Mpc$^{-1}$. 

\begin{table}
\begin{center}
\caption{Summary of the spectral fitting of the \emph{Fermi}/LAT data integrated on long timescales. $F_{E>100 \rm MeV}$ is in units of [$10^{-6}$~ph~cm$^{-2}$~s$^{-1}$]. The statistical measurement errors are at $1\sigma$. Systematic errors are not included. The latest estimates for the adopted Instrument Response Function (IRF) \textsc{P6\_V3\_DIFFUSE} are: 10\% at 100 MeV, 5\% at 500 MeV and 20\% at 10 GeV, while the error in the photon index is $\sim 0.1$ (Rando et al. 2009).}
\begin{tabular}{lccc}
\hline
Time Period & $F_{E>100 \rm MeV}$ & $\Gamma$ & TS\\
\hline
Jun $01-30$ & 0.23$\pm$0.01   & 2.77$\pm$0.06 & 98 \\
Jul $07-10$ & 1.02$\pm$0.02   & 2.55$\pm$0.02 &  140 \\
Aug 1 $-$ Sep 14   & 0.26$\pm$0.01 & 2.74$\pm$0.03 & 140 \\
\hline
\end{tabular}
\end{center}
\label{lat}
\end{table}

\begin{table*}
\begin{center}
\caption{Summary of results from analysis of the \emph{Swift} data obtained on 2010~July~3 (ObsID~00038394002) compared with the observation of 2008~December~5 (ObsID~00031306001, Discovery 2008) and 2009~May~15 (ObsID~00031306006, MW Campaign 2009), both reanalyzed here with the latest software (with consistent results), when the source was at low $\gamma$-ray flux. The value of $N_H$ was fixed to the Galactic column as measured by Kalberla et al. (2005).}
\begin{tabular}{ccccccc}
\hline
\multicolumn{7}{c}{XRT ($0.3-10$~keV)}\\
\hline
ObsID & Date & Exp. & $N_H$ & $\Gamma$ & Flux$_{0.3-10 \rm keV}$ & $\chi^{2}$/dof \\
{} & [YYYY/MM/DD] & [ks] & [$10^{20}$~cm$^{-2}$] & {} & [$10^{-12}$~erg~cm$^{-2}$~s$^{-1}$] & {} \\
\hline
00038394002 & 2010/07/03 & 1.6 & 5.22 & 1.5$\pm$0.3 & 5.5$\pm$0.5 & 0.53/2 \\
00038306006 & 2009/05/15 & 1.3 & 5.22 & 1.7$\pm$0.3 & 2.3$\pm$0.3 & ($^*$)\\ 
00031306001 & 2008/12/05 & 4.2 & 5.22 & 1.8$\pm$0.2 & 3.9$\pm$0.2	& 3.2/8 \\
\hline
\multicolumn{7}{l}{$^*$ \scriptsize{Low statistics. The fit has been performed by using the maximum likelihood (Cash 1979), obtaining c-stat of 57 for 58 d.o.f.}}\\
\hline
\multicolumn{7}{c}{UVOT (observed magnitudes)}\\
\hline
ObsID & $v$ & $b$ & $u$ & $uvw1$ & $uvm2$ & $uvw2$ \\
\hline
00038394002 & 17.6$\pm$0.2 & 18.3$\pm$0.1 & 17.24$\pm$0.08 & 17.19$\pm$0.07 & 17.34$\pm$0.09 & 17.23$\pm$0.06\\
00031306006 & 18.3$\pm$0.2 & 18.4$\pm$0.1 & 17.7$\pm$0.1 & 17.43$\pm$0.08 & 17.5$\pm$0.1 & 17.65$\pm$0.07\\
00031306001 & 18.2$\pm$0.1 & 18.56$\pm$0.08 & 17.79$\pm$0.07 & 17.57$\pm$0.06 & 17.63$\pm$0.07 & 17.63$\pm$0.05\\
\hline
\end{tabular}
\end{center}
\label{swift}
\end{table*}

\begin{figure}
\begin{center}
 \includegraphics[angle=270,scale=0.35]{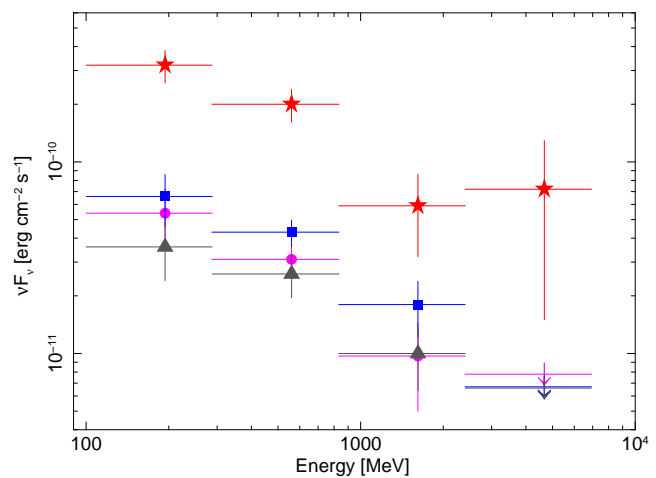} 
 \caption{High-energy $\gamma$-ray SED of PMN~J0948$+$0022. The red star symbols refer to the period of maximum flux (2010~July~7$-$10, 4 days), the blue squares indicate the emission integrated over the period 2010~August~1~$-$~September~15 (45 days), while the magenta circles are for the period 2010~June~1$-$30 (30 days). The grey triangles refer to the quiescence period of one month centered on 2008~December~5 (corresponding to the \emph{Swift} observation, Abdo et al. 2009b). All the detections have $TS>10$ ($\sim 3\sigma$, see Mattox et al. 1996). Upper limits are at $2\sigma$ level. See also Table~1.}
   \label{fig:sed}
\end{center}
\end{figure}

\section{Data Analysis}
The analysis of the data from the different instruments and facilities participating in the present campaign was done in basically the same way as for that of 2009 (see MW Campaign 2009 paper and references therein). There were some updates in the software used: \emph{Swift}/XRT and UVOT data were studied with the specific packages included in \textsc{HEASoft v. 6.10}\footnote{Including the XRT Data Analysis Software (XRTDAS) developed under the responsibility of the ASI Science Data Center (ASDC), Italy.} and the calibration database updated on 2010~September~30. 

The light curve obtained from \emph{Fermi}/LAT data is shown in Fig.~\ref{fig1}, while the spectra are displayed in Fig.~\ref{fig:sed} and the fits are summarized in Table~1. The change in the photon index, confirmed by the spectra shown in Fig.~\ref{fig:sed}, indicates some ``harder when brighter'' behavior. This comparative analysis should be considered with care and the caveats about the comparison of fits over the whole energy band versus fits on separate energy bins are described in detail in the Section 4.4 of Abdo et al. (2010a). 

Fig.~\ref{fig2} shows the optical (R filter) and radio (15~GHz) light curves from ATOM (\emph{Automatic Telescope for Optical Monitoring for H.E.S.S.}, Namibia) and OVRO (\emph{Owens Valley Radio Observatory}) 40~m telescope, the latter as part of an ongoing \emph{Fermi} blazar monitoring program (Richards et al. 2009). Multifrequency observations to measure spectral changes were also done at Effelsberg (Project F-GAMMA, Fuhrmann et al. 2007) and RATAN-600 (Kovalev et al. 1999), respectively. The study of the morphology was done with Very Long Baseline Interferometry (VLBI) observations performed on 2010~September~17 as part of the MOJAVE monitoring program\footnote{\texttt{http://www.physics.purdue.edu/astro/MOJAVE/}} conducted with the Very Long Baseline Array (VLBA) at $\lambda = 2$~cm and provided high resolution total intensity and linear polarization images. Detailed description of the MOJAVE program as well as method of observations, data processing and analysis is presented by Lister et al. (2009). 

The source was close to the Sun in 2010~July and therefore it was not possible to perform a densely-sampled optical-to-X-ray follow-up. Only one X-ray observation close to the outburst remained and was performed by \emph{Swift} on July~3 (ObsID~00038394002\footnote{Within the project \emph{Swift-XRT Monitoring of Fermi-LAT Sources of Interest}. See: \texttt{http://www.swift.psu.edu/monitoring/}.}), about 4 days before the outburst. A summary of XRT and UVOT analysis is displayed in Table~1, compared with the observation of 2008~December~5 (ObsID~00031306001, Discovery 2008) and of 2009~May~15 (ObsID~00031306006, the lowest X-ray flux measured during the MW Campaign 2009), when the source was in a low activity state at $\gamma$ rays. 

\begin{figure}
\begin{center}
 \includegraphics[scale=0.85]{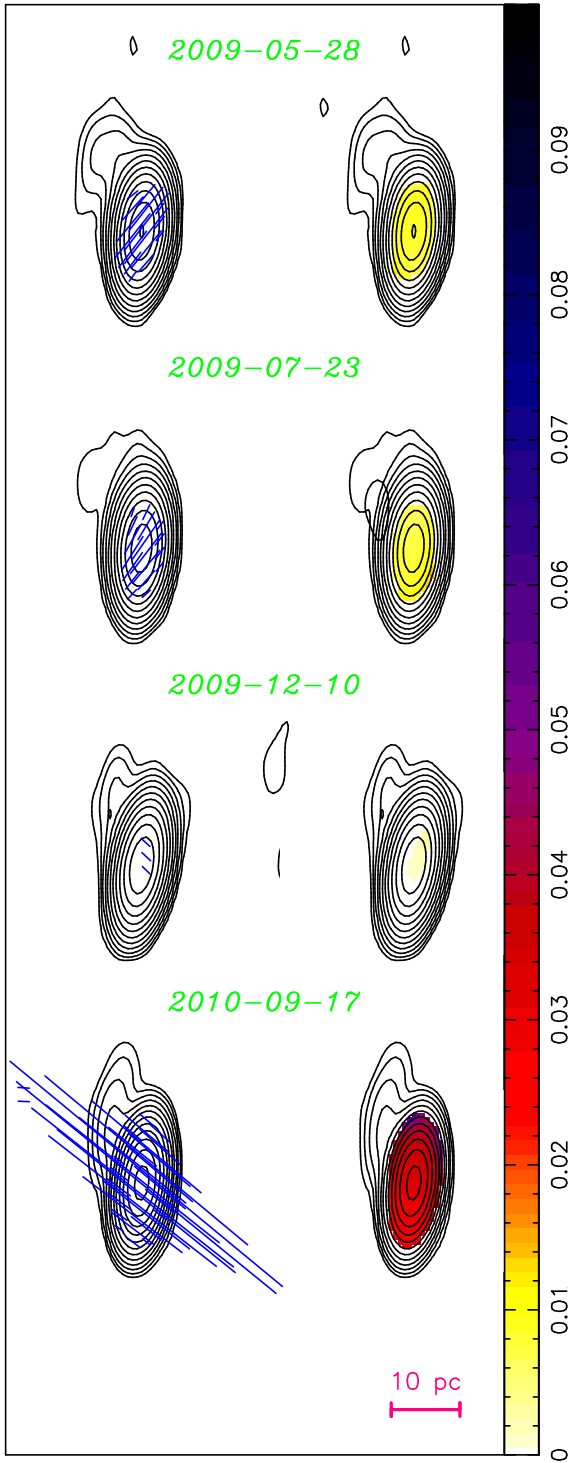} 
 \caption{Total intensity and linear polarization images observed by VLBA at 15~GHz in different epochs as part of the MOJAVE program. Naturally-weighted total intensity images are shown by black contours, the contours are in successive powers of two times the base contour level of 0.2~mJy~beam$^{-1}$. Electric polarization vectors direction is indicated on the left hand side by blue sticks, their length is proportional to the polarized intensity. Linear fractional polarization is shown on the right hand side overlaid according to the color wedge.}
   \label{fig:mojave}
\end{center}
\end{figure}

\begin{figure}
\begin{center}
 \includegraphics[angle=270,scale=0.35]{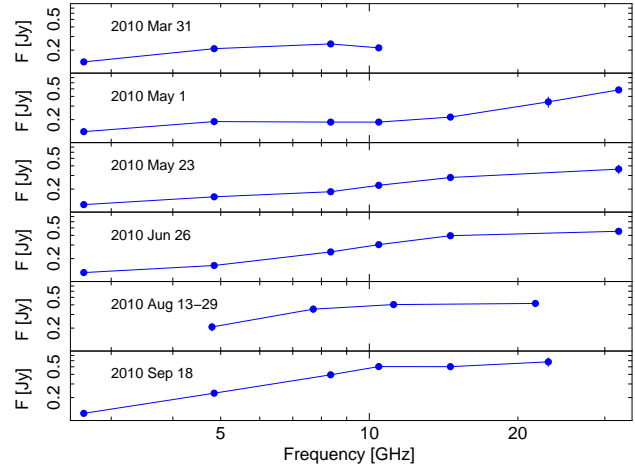} 
 \caption{Spectral evolution at radio frequencies as measured by RATAN-600 (2010~August~13$-$29) and Effelsberg (all the remaining measurements).}
   \label{fig:radiospec}
\end{center}
\end{figure}

The data available confirm the outburst first detected at $\gamma$ rays of the NLS1 PMN~J0948$+$0022. The optical light curve from ATOM (R filter) indicates an increase of the optical activity with a change of about one magnitude during the early six months of 2010 (Fig.~\ref{fig2}, \emph{top panel}). This trend is confirmed by the \emph{Swift}/UVOT observation performed on July~3, when compared with an observation of the source at low $\gamma$-ray state (2008~December~5 in the Discovery 2008; 2009~May~15 in the MW Campaign 2009): the decrease of the observed magnitude ranges from 0.6$-$0.7 in the $v$ filter to 0.3$-$0.4 magnitudes in the ultraviolet filters. The same behavior was observed during the MW Campaign of 2009~March-July, when the source flux was decreasing, but showing stronger variability at optical frequencies rather than at the ultraviolet ones. This is likely to be due to the fact that ultraviolet frequencies sample the emission from the accretion disk, while the optical emission is mainly due to the synchrotron radiation. 

The X-ray flux as measured by \emph{Swift}/XRT was $\sim 40$\% greater than the value measured in the Discovery 2008 paper and a factor $\sim 2.4$ greater with respect to the lowest flux measured during the MW Campaign 2009. The low statistics in the short observations of 2010~July~3 and 2009~May~15 prevent any conclusion on possible spectral changes. 

At $\gamma$ rays there was the most dramatic change in the emission of the electromagnetic radiation with an increase of flux by a factor $\sim 5$ in the 0.1$-$100~GeV energy band. The outburst\footnote{Some information were derived from the analysis of the LAT light curve with 1 day time bin not shown here.} (Fig.~\ref{fig1}) was characterized by a first sharp increase of the flux on July~7 (Donato et al. 2010) with $e-$folding timescale\footnote{That is the characteristic time $\tau$ to have an exponential increase or decrease of flux: $F(t)=F(t_0)\exp[-(t-t_0)/\tau]$.} of 3.6$\pm$1.8 days, reaching the peak value of $\sim 10^{-6}$~ph~cm$^{-2}$~s$^{-1}$ and remaining at such level for about four days. On July~11 the source was not detected (with a $2\sigma$ upper limit of $2.7\times 10^{-7}$~ph~cm$^{-2}$~s$^{-1}$) and then it was detected again on July~12 above $\sim 10^{-6}$~ph~cm$^{-2}$~s$^{-1}$ ($e-$folding rise time $<1$~day). After a few days of low activity, there was a second shorter outburst (Foschini 2010) with flux $\sim 10^{-6}$~ph~cm$^{-2}$~s$^{-1}$ and rise time $<3$~days. Then the source slowly returned to quiescence, with some occasional detection at moderate flux. 

The spectral behavior is more difficult to study, as previously outlined. Nevertheless, the comparative analysis based on long timescale (month) integration, thus with a more robust statistics than that available on short timescales (Fig.~\ref{fig:sed}, Table~1), indicates that there is some spectral hardening between the spectrum measured during the four days of the peak and the spectra measured on the data integrated on 2010~June and August-September. It is not possible to firmly establish if this is due to a change of the peak of the inverse-Compton emission, as occurred in the case of 3C~454.3 (see Bonnoli et al. 2011), or if it is due to an effective increase of the population of GeV photons. 

The spectral evolution of the radio emission, as measured by Effelsberg and RATAN-600 (Fig.~\ref{fig:radiospec}) indicate instead a progressive inversion of the spectral index between 4.8 and 10 GHz ($S_{\nu}\propto \nu^{-\alpha}$), changing from $\alpha=0.02\pm 0.03$ on 2010~May~1 (MJD~55317.83) to $\alpha=-0.81\pm 0.02$ on 2010~June~26 (MJD~55373.77), an average $\alpha=-0.80\pm 0.21$ measured in the period 2010~August~13$-$29, and up to a value of $\alpha = -1.0\pm 0.1$ measured on 2010~September~18 (MJD~55457.45). 

We compare these results with the MW Campaign of 2009~March-July, reported in Abdo et al. (2009c). In that case, the source was at a moderate high flux in 2009~April, peaking on April~1\footnote{As indicated in the report \#44 (2009~April~1$-$7) of the \emph{Fermi Gamma ray Sky} blog \texttt{http://fermisky.blogspot.com/}.} with a flux of $(4.2\pm 0.9)\times 10^{-7}$~ph~cm$^{-2}$~s$^{-1}$ in the 0.1$-$100~GeV energy band and $\Gamma = 2.1\pm 0.1$ ($TS=46$; reanalyzed in the present work). However, the flux integrated over the whole month of April gave a softer spectrum ($\Gamma = 2.7\pm 0.2$, cf Abdo et al. 2009c). In the 2010~July outburst, the $\gamma$-ray spectrum seems softer before (June) and after (August-September) the burst, with some hardening during the peak of the emission (July). The 2010~July burst was a factor of $\sim 2$ greater than the 2009~April flare and also longer (more than 4 days vs 1 day). 

\begin{figure*}
\begin{center}
 \includegraphics[scale=0.47,clip,trim=0 40 50 80]{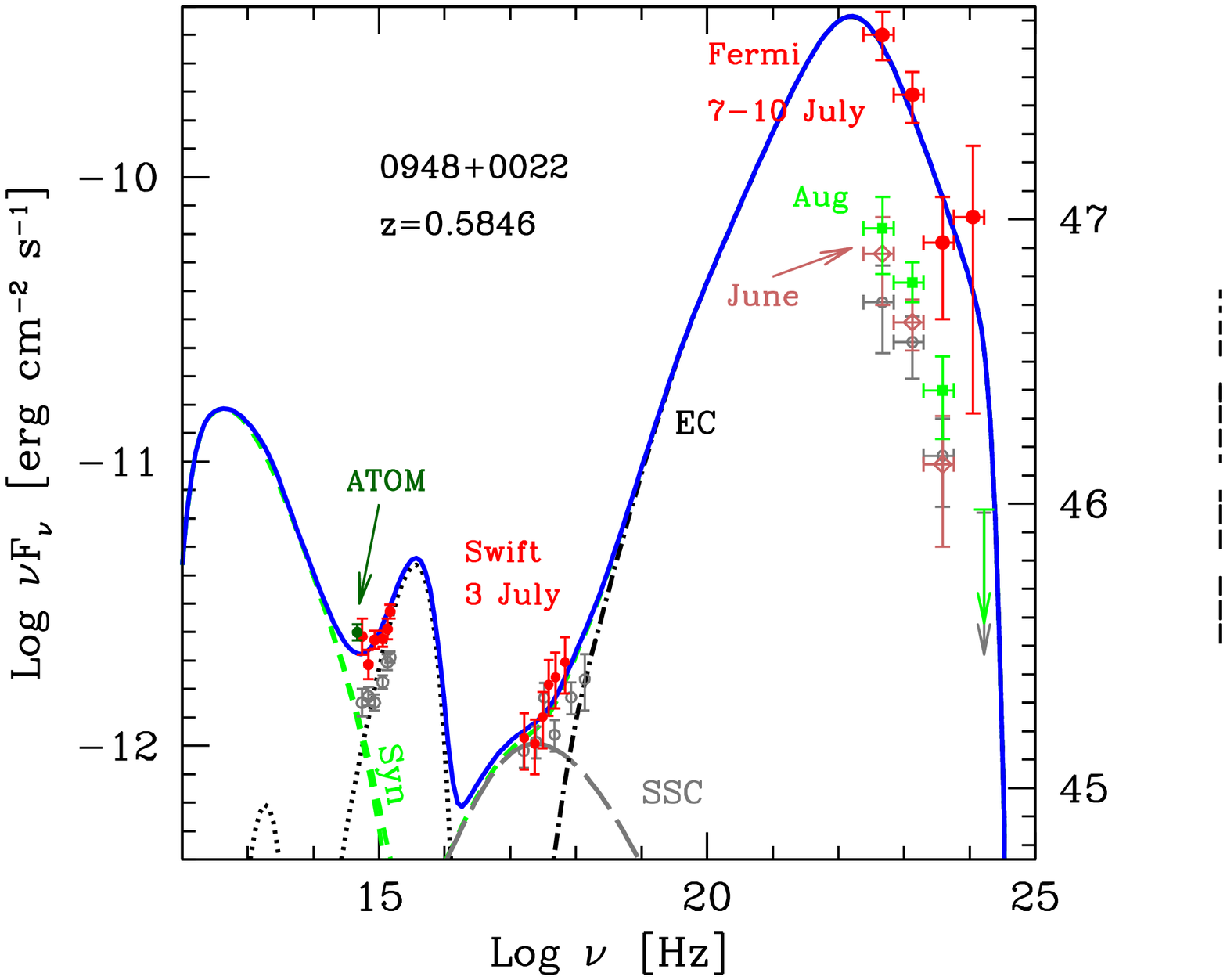}
  \includegraphics[scale=0.47,clip,trim=50 40 0 40]{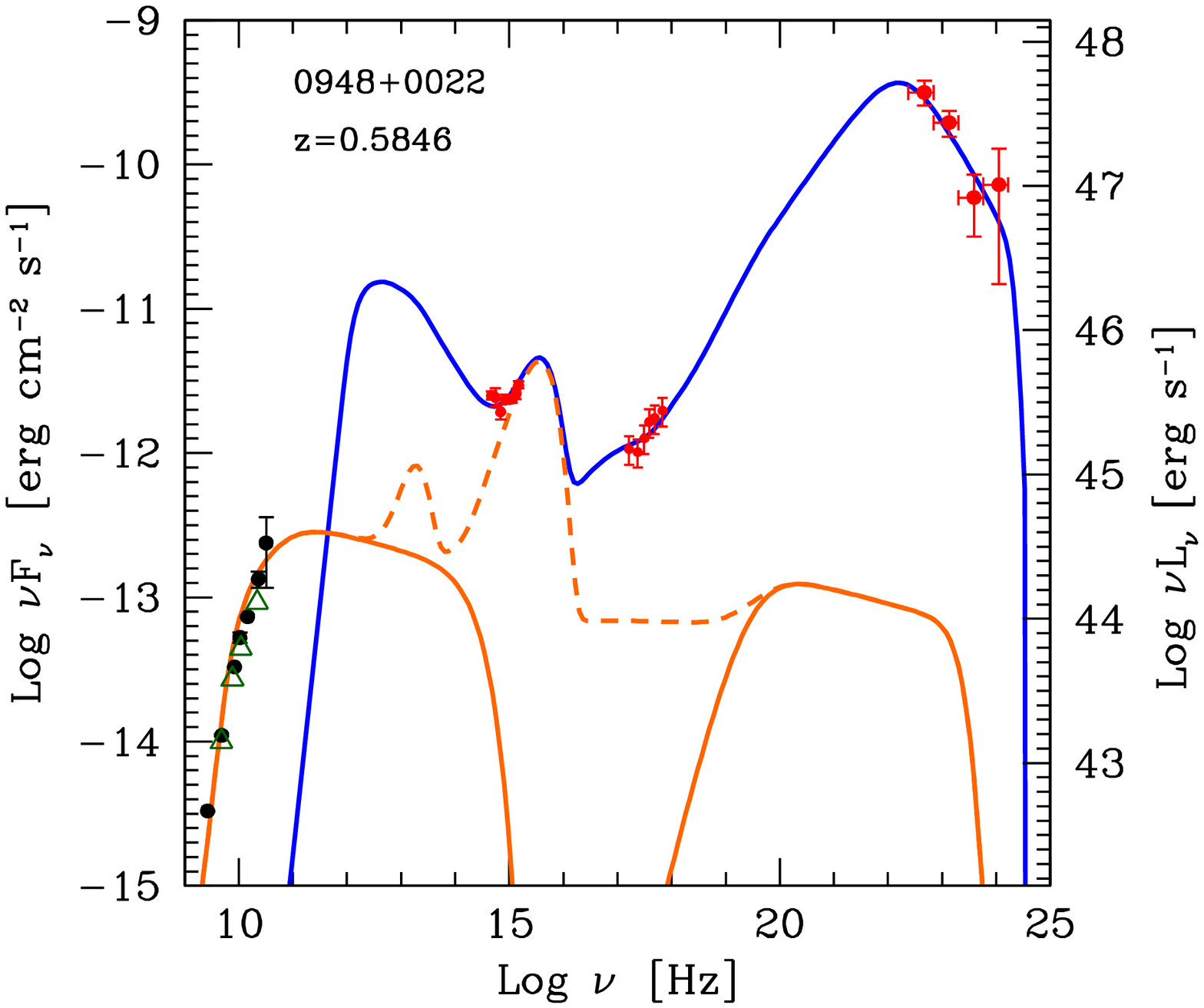} 
 \caption{(\emph{left panel}) SED of PMN~J0948$+$0022. The grey symbols refer to the low activity period observed by \emph{Swift} (5~December~2008) and \emph{Fermi} (one month centered on the \emph{Swift} observation). The red symbols, used to model the SED, refer to July~7$-$10 (\emph{Fermi}/LAT), July~3 (\emph{Swift} XRT and UVOT), and June~26 (ATOM R filter). The model components are: synchrotron (green dashed line), accretion disk and torus (black dotted line), synchrotron self-Compton (grey long-dashed line), external Compton (black dot-dashed line) and the sum of all of them (blue continuous line). (\emph{right panel}) SED with radio data from RATAN-600 (August~13$-$29, green open triangles) and Effelsberg (September~18, black filled circles). The orange continuous lines are the result of the modeling of radio data (synchrotron plus external Compton with seed photons from the molecular torus). The dashed orange line is the emission from the torus and the accretion disk. The blue continuous line and the red points are the SED of the peak (July~7$-$10) shown in the \emph{left panel}.}
   \label{fig:seds}
\end{center}
\end{figure*}

It is also interesting to note the evolution of the radio emission, polarization and morphology as shown by the VLBA observations at 15~GHz (Fig.~\ref{fig:mojave}). The observation performed on 2010~September~17 shows a bright unresolved core with total parsec-scale flux density as high as 538~mJy and a high degree of polarization (3.3\%), greater than the values measured in 2009 ($<1$\%). Specifically, when studying the evolution from 2009~May (MW Campaign 2009) to 2010~September, we noted a drop in the polarized flux density in the third epoch, coupled with a swing of the electric vector polarization angle (EVPA) of about 90$^{\circ}$, which is maintained in the last epoch, even with a much greater polarized flux density (a factor $\sim 6$). A similar behavior has been observed -- for example -- before the 2008~August outburst at $\gamma$ rays of the flat-spectrum radio quasar PKS~1502$+$106 ($z=1.839$, Abdo et al. 2010b).

It is not possible to estimate the speed of the radio jet on the basis of these 4 epochs only, particularly as in the present case when the source is strikingly compact. Observations are continuing, and we anticipate having a robust speed estimate in the near future using a larger sample of epochs.

\section{Modeling of the SED}
The spectral energy distribution shown in Fig.~\ref{fig:seds} (\emph{left panel}) has been fitted with the model described in Ghisellini \& Tavecchio (2009), the same used in the previous works (Abdo et al. 2009a,b,c). The electromagnetic emission from a relativistic jet is modeled with different components: the synchrotron radiation plus the inverse Compton emission derived by the interaction of the relativistic electrons with different types of seed photons (synchrotron, molecular torus, broad-line region, accretion disk, corona). We present here our favored interpretation of the collected data. Changes in the model parameters are negligible with respect to variations of $\pm 30$\% in the 0.1$-$100~GeV flux or $\pm 0.1$ in the photon index (see Section 4 of Discovery 2008 paper for a more extended discussion on the impact of measurement errors on the calculations of the model parameters). 

The $\gamma$-ray flux is most important, because it flags the jet power. The extreme Compton dominance suggests a Doppler factor greater than that measured during the MW Campaign 2009: therefore, we adopted in the model a bulk Lorentz factor $\Gamma=16$ (vs 10 of the MW Campaign 2009), with a jet viewing angle of $3^{\circ}$ (vs $6^{\circ}$). This is one ``economical'' possibility, in terms of energetic. It would be possible to obtain the same Compton dominance with a lower magnetic field, but with greater injected power (see below). In addition, it is worth noting that the magnetic field is linked to the X-ray emission through the synchrotron self-Compton process (it is peaking at these energies) and the only available X-ray measurement has been performed well before the beginning of the outburst (Table~2). 

The energy distribution of the injected electrons (with power $2.1\times 10^{43}$~erg~s$^{-1}$ in the comoving frame) spans from $\gamma_{\rm e,min}=1$ to $\gamma_{\rm e,max}=2700$ and is described by a broken power-law model with index $\gamma_{\rm e}^{2}$ and $\gamma_{\rm e}^{-3.2}$ below and above the break $\gamma_{\rm e,break}=200$, respectively, where $\gamma_{\rm e}$ indicates the random Lorentz factor of the electrons. The external Compton peak is produced by electrons with $\gamma_{\rm e,peak} \simeq 170$, scattering off seed photons of the broad-line region. 

Our model shows that most of the dissipation occurs at about $1.2\times 10^{17}$~cm ($2600$ Schwarzschild radii, within the broad-line region), the magnetic field is $B=2$~G. The mass of the central black hole was fixed to the value of $1.5\times 10^{8}M_{\odot}$, as measured by fitting of the accretion disk emission (see Discovery 2008) and in agreement with the measurement by Zhou et al. (2003) by using FWHM of emission lines. The accretion rate is at $50$\% of the Eddington value. The calculated jet powers (the output of the model) are $P_{\rm radiative} = 5\times 10^{45}$, $P_{\rm magnetic} = 1.2\times10^{45}$, $P_{\rm electrons} = 6\times 10^{44}$ and $P_{\rm protons} = 10^{47}$~erg~s$^{-1}$, by assuming one proton per electron. 

These values have to be compared with those reported in previous studies (specifically, see Table 4 of MW Campaign 2009 paper). The present values of the jet powers are generally greater, up to one order of magnitude in the case of the protons. Other differences with respect to the previous works are a lower magnetic field with respect to the average of the MW Campaign 2009 ($B=4.1$~G), but similar to the value in Discovery~2008 ($B=2.4$~G). The injected power is similar to the previous studies, but the shape is different, with a greater $\gamma_{\rm e,max}$ (2700 vs 2000 and 1600, in the MW Campaign 2009 and Discovery 2008, respectively) and a smaller $\gamma_{\rm e,break}$ (200 vs 530 and 800, in the MW Campaign 2009 and Discovery 2008, respectively). 

Since optical-to-X-ray data have been measured a few days before the outburst, it is possible that they are underestimated. If we assume an increase by a factor 2-4 in the optical-to-X-ray fluxes, keeping the $\gamma$-ray flux to the observed peak value, then it is possible to model the SED with a relatively smaller bulk Lorentz factor ($\Gamma=13$). The smaller beaming is compensated by an increase of the injected power of about one third, plus a greater magnetic field ($B=3.6$~G). The jet powers (protons, electrons, ...) have negligible changes, because they are strongly dependent on the $\gamma$-ray flux.

The region emitting the optical-to-$\gamma$-ray flux is too compact to be responsible for the radio emission (its synchrotron flux is self-absorbed). This is confirmed by the variability analysis: the timescales at $\gamma$ rays are short, of the order of days or even less than one day, thus requiring a very compact source. About two months after the outburst at $\gamma$ rays, the radio flux density reached its maximum (Fig.~\ref{fig2}, \emph{bottom panel}). During this period, the emitting region can move outward and expand, thus becoming optically thin at radio frequencies. Moreover, near-simultaneous OVRO and MOJAVE observations indicate $\la 50$~mJy of arcsecond-scale flux density at 15~GHz, hence nearly all of the radio emission in the source is generated on pc-scales. Therefore, to account for the radio emission, we considered an additional larger emitting zone (Fig.~\ref{fig:seds}, \emph{right panel}). For the sake of simplicity, we assume the same bulk Lorentz factor of the optical-to-$\gamma$ rays fit ($\Gamma = 16$). This region is $\sim 0.6$~pc in size, about two orders of magnitude larger than the $\gamma$-ray emitting region, and farther (5.8~pc) from the central black hole. At this distance, the seed photons for the external Compton process are from the molecular torus, emitting at infrared wavelengths. 

\section{Discussion and Conclusion}
We presented the results of a MW Campaign from radio to $\gamma$ rays performed in 2010 to study the evolution of the electromagnetic emission of the Narrow-Line Seyfert 1 Galaxy PMN~J0948$+$0022.

\begin{figure}
\begin{center}
 \includegraphics[scale=0.4,clip,trim=0 40 0 40]{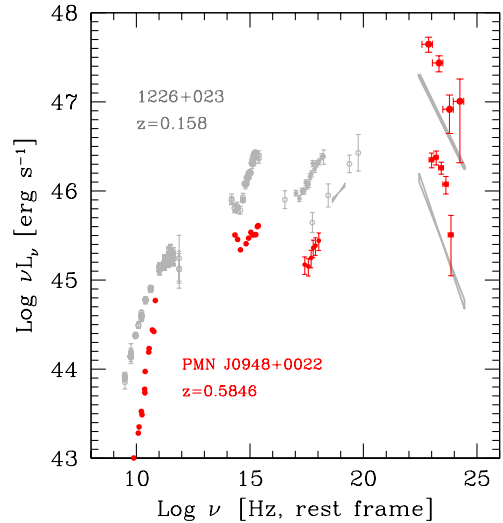} 
 \caption{Comparison of the SEDs of the archetypical blazar 1226$+$023 (3C~273, grey symbols) and the NLS1 PMN~J0948$+$0022 (red symbols). In the case of high-energy $\gamma$ rays, the minimum and maximum flux detected by \emph{Fermi} are shown.}
   \label{fig:3C273}
\end{center}
\end{figure}

Despite all the caveats and uncertainties, what clearly emerges from this episode is the ability of a NLS1 to generate an extreme power ($\sim 10^{48}$~erg~s$^{-1}$), when compared to that of a typical flat-spectrum radio quasar. Fig.~\ref{fig:3C273} shows a comparison of the SED (rest frame) of PMN~J0948$+$0022 with that of the archetypical blazar 3C~273, known to have a strong disk emission too. At first look, the figure shows that the NLS1 has a more extreme Compton dominance: although its radio-to-X-ray luminosity is smaller than that of 3C~273, the $\gamma$-ray power is greater. If we renormalize the two SEDs by taking as reference the peak due to the accretion disk emission, then it is necessary to multiply the values of PMN~J0948$+$0022 by a factor $\sim 6$, which in turn is roughly similar to the ratio of the masses of the central black holes ($1.5\times 10^{8}M_{\odot}$ for PMN~J0948$+$0022 and $8\times 10^{8}M_{\odot}$ for 3C~273). By shifting upward the SED of PMN~J0948$+$0022 by this value, the differences in the radio-to-X-ray emission become negligible, while those at $\gamma$ rays are emphasized. At these energies, the differences can be explained with the different Doppler factor as a consequence of the different viewing angle (smaller in the case of this NLS1).

The basic findings of the present work can be summarized as follows:

\begin{itemize}
\item The source reached an observed luminosity of $\sim 10^{48}$~erg~s$^{-1}$ at $\gamma$ rays (0.1$-$100~GeV), with day-scale variability and some ``harder when brighter'' spectral behavior. This is the first time that such a power is measured from a NLS1. It confirms, beyond any reasonable doubt, that NLS1s can host relativistic jets as powerful as those in blazars and radio galaxies, despite the relatively low mass ($1.5\times 10^{8}M_{\odot}$) and the rich environment due to a high accretion rate (50\% the Eddington rate). 

\item The coordinated variability observed at all the available frequencies confirms the typical behavior of relativistic jets, as already shown during the MW Campaign 2009. Particularly, the radio emission peaked about two months after the outburst at $\gamma$ rays, although the radio spectral index was already inverted a couple of weeks before the peak in the GeV band. The optical flux changed about one magnitude (R filter) during the months before the outburst.

\item The morphology at 15~GHz still shows a very compact source ($\sim 10$~pc size), despite of the great power released at $\gamma$ rays. Comparison with previous observations in 2009 shows that the EVPA changed $\sim 90^{\circ}$ at some time between 2009 July and December and this new value is maintained in 2010. The linear polarization fraction exceeds 3\%, a value greater than that measured in 2009 ($<1$\%).  
 
\item The comparison with the SED of a typical blazar with a strong accretion disk (3C~273) shows that the Compton dominance is more extreme in the NLS1. The disagreement of the two SEDs can be accounted by the differences in mass of the central black hole and Doppler factor of the two jets.
\end{itemize}

\section*{Acknowledgments}
This research has made use of data obtained from the High Energy Astrophysics Science Archive Research Center (HEASARC), provided by NASA's Goddard Space Flight Center. 

This work has been partially supported by PRIN-MiUR 2007 and ASI Grant I/088/06/0. 

The OVRO 40~m monitoring program is supported in part by NASA (NNX08AW31G) and the NSF (AST-0808050). 

Based on observations with the 100-m telescope of the MPIfR (Max-Planck-Institut f\"ur Radioastronomie) at Effelsberg. Ioannis Nestoras is member of the International Max Planck Research School (IMPRS) for Astronomy and Astrophysics at the Universities of Bonn and Cologne. 

The \emph{Fermi} LAT Collaboration acknowledges support from a number of agencies and institutes for both development and the operation of the LAT as well as scientific data analysis. These include NASA and DOE in the United States, CEA/Irfu and IN2P3/CNRS in France, ASI and INFN in Italy, MEXT, KEK, and JAXA in Japan, and the K.~A.~Wallenberg Foundation, the Swedish Research Council and the National Space Board in Sweden. Additional support from INAF in Italy and CNES in France for science analysis during the operations phase is also gratefully acknowledged. 

This research has made use of data from the MOJAVE database that is maintained by the MOJAVE team (Lister et al. 2009). The MOJAVE project is supported under NSF grant AST-0807860 and NASA {\it Fermi} grant NNX08AV67G. 

RATAN-600 observations were supported in part by the Russian Foundation for Basic Research grant 08-02-00545. Y.~Y.~Kovalev was supported in part by the return fellowship of Alexander von Humboldt Foundation.

\label{lastpage}

\end{document}